\documentclass{aa}  

\usepackage{graphicx}
\usepackage{txfonts}
\usepackage[colorlinks=true,citecolor=blue]{hyperref}
\usepackage{placeins}

\usepackage{booktabs}
\usepackage{multirow}
\usepackage[dvipsnames,table]{xcolor}

\usepackage{nicefrac}

\usepackage{siunitx}
\DeclareSIUnit\Lsun{L\ensuremath{_\odot}}
	\DeclareSIUnit\Msun{M\ensuremath{_\odot}}
	\DeclareSIUnit\Rsun{R\ensuremath{_\odot}}
	\DeclareSIUnit\parsec{pc}
        \DeclareSIUnit\jansky{Jy}
        \DeclareSIUnit\magnitude{mag}

\DeclareMathOperator\erf{erf}
\newcommand{\PA}{\mathit{P\!A}} 

\usepackage{pifont}
\newcommand{\cmark}{\ding{51}}
\newcommand{\xmark}{\ding{55}}

\definecolor{bubblegumpink}{HTML}{FE83CC}
\definecolor{bluegreen}{HTML}{137E6D}
\definecolor{cobaltblue}{HTML}{030AA7}
\definecolor{skyblue}{HTML}{448EE4}
\definecolor{scarlet}{HTML}{BE0119}

\definecolor{lightforestgreen}{HTML}{4F9153}
\definecolor{darkishgreen}{HTML}{287C37}
\definecolor{mutedgreen}{HTML}{5FA052}
\definecolor{lawngreen}{HTML}{4DA409}
\definecolor{kelleygreen}{HTML}{009337}
\definecolor{darkgrassgreen}{HTML}{388004}

\begin{document}

   \title{Spatial distribution of water ice in the protoplanetary silhouette disk d216-0939}

   \subtitle{}

   \author{
     J.~S.~Martin\inst{1}
     \and
     S.~Wolf\inst{1}
     \and
     A.~Potapov\inst{2}
     \and
     H.~Linz\inst{3}
     \and
     Th.~Henning\inst{3}
     \and
     H.~Terada\inst{4}
     \and
     M.~B.~Michaelis\inst{1}
   }

   \institute{
     Institute of Theoretical Physics and Astrophysics,
     Kiel University,
     Leibnitzstraße 15, 24118 Kiel, Germany\\
     \email{jmartin@astrophysik.uni-kiel.de}
     \and
     Analytical Mineralogy Group,
     Institute of Geosciences,
     Friedrich Schiller University Jena,
     CEEC II, Lessingstr. 14, 07743 Jena, Germany
     \and
     Max Planck Institute for Astronomy, Königstuhl 17, 69117 Heidelberg, Germany
     \and
     National Astronomical Observatory of Japan, Tokyo, Japan
   }

   \date{\today}
 
   \abstract
   {The composition of rocky planets depends on the dust in their natal protoplanetary disk (PPD), potentially containing water ice. Crystalline water ice was detected in the PPD d216-0939 in the Orion Nebular Cluster (ONC).} 
   {We aim at constraining the spatial distribution and crystallization state of water ice in the d216-0939 disk using recent observations of the water ice absorption feature at a wavelength of $\sim$\SI{3}{\micro\meter} collected with JWST.}
   {We perform 3D Monte Carlo radiative transfer (MCRT) simulations to constrain the free parameters of an accretion disk model by fitting the calculated spectrum in the wavelength range from \SIrange{1.6}{24}{\micro\meter} to the spectral energy distribution (SED) observed with the JWST instruments NIRSpec and MIRI. Additionally, archival, spatially resolved HST observations were used to constrain the global spatial structure of the disk, as the parameter space is degenerate with respect to the fit to the SED. Successively, we fit the water ice absorption feature using polychromatic MCRT simulations with spectral importance sampling to produce synthetic observations at high spectral resolutions.}
   {By probing the upper and outer disk layers, we found that a PPD model with dust containing \SI{5.4}{\percent} crystallized water ice beyond the snowline fits the observations well, necessitating outward transport of material, since crystalline ice is unlikely to form in situ in the probed disk layers. In the spectral region of the water ice absorption feature, scattering and thermal dust emission both contribute significantly to the total flux.}
   {}


   \maketitle
   
   \section{Introduction}
   Observations of protoplanetary disks (PPDs), the progenitors of planetary systems, allow us to study the physical conditions in which planet formation takes place \citep[e.g.,][]{Keppler+2018,vanCapelleveen+2025}. The presence of ices in PPDs impacts the process of planet formation, as their freeze-out from the gas phase defines snowlines in the disks \citep[e.g.,][]{Oeberg+2011,Leemker+2026}. Solid-state water is the most abundant ice in astrophysical environments \citep[e.g.,][]{vanDishoeck+2021,Ballering+2021} and its spatial distribution in PPDs is linked to the water delivery process to forming planets \citep[e.g.,][]{Sato+2016,Topchieva+2024,Williams+2025}. For Earth, this process is not yet conclusively explained by theories on the formation and evolution of the Solar System \citep[e.g.,][]{Hartmann+2017,OBrian+2018}.
   
   Water ice was found in many PPDs \citep[e.g.,][]{SchegererWolf2010,Aikawa+2012,Terada+2012,TeradaTokunaga2017}, with an almost edge-on geometry favoring the detection via the stretching mode of H\textsubscript{2}O ice in transmission with a minimum at a wavelength of $\sim$\SI{3}{\micro\meter} (``water ice feature''). Observations with the \textit{James Webb Space Telescope} (JWST) in the infrared (IR) reveal complex chemistry of ices in PPDs \citep[e.g.,][]{Rocha+2024b,Potapov+2025}. The shape of the water ice feature is not only influenced by the chemical composition and crystallinity of the ice but additionally depends on the grain size and disk geometry \citep[e.g.,][]{Pontoppidan+2005,Ballering+2021,Sturm+2023b,Dartois+2024,Martinien+2025a}.
   
   Detailed studies of the PPDs HH 48 NE \citep{Sturm+2023a,Sturm+2023b,Sturm+2023c,Bergner+2024}, PDS 453 \citep{Martinien+2024}, Tau 042021 \citep{Dartois+2025}, and d114-426 \citep{Ballering+2025} demonstrate the usefulness of 3D radiative transfer (RT) simulations to draw conclusions on the physical origin of the water ice feature. These studies show that the spatial structure of PPDs, i.e., large inner cavities in HH 48 NE and PDS 453, a tilted inner disk in d114-426, and a disk wind in Tau 042021, influences the appearance of edge-on disks. Additionally, scattered radiation shifts the minimum of ice features \citep{Martinien+2025a}, may saturate them \citep{Martinien+2025b}, and, depending on the inclination angle of the disk, may lead to excess emission in the ice features \citep[e.g.,][]{Dartois+2022}.
   
   The silhouette disk d216-0939 in the Orion Nebular Cluster (ONC), which we investigate in this work, was observed with the \emph{Hubble Space Telescope} \citep[HST;][]{Smith+2005}. Here, the shadow of the PPD is seen against the H$\alpha$ background emission of the outskirts of the ONC. Observations of the disk have been obtained at (sub-)millimeter wavelengths with the Submillimeter Array \citep[SMA;][]{MannWilliams2009}, the Very Large Array \citep[VLA;][]{Sheehan+2016}, and, finally, the Atacama Large Millimeter/submillimeter Array \citep[ALMA;][]{Mann+2014,Factor+2017,Sheehan+2020,Diaz-Berr+2024}. \citet{Sheehan+2020} report that d216-0939 (alternatively named HOPS-65) has dispersed its natal envelope and a \SI{216}{\astronomicalunit} wide gap in the large grain population (up to approximately \SI{50}{\milli\meter}) centered at a radial distance of \SI{111}{\astronomicalunit} from the protostar is revealed by spatially resolved observations. Photometry in IR to submillimeter wavelengths collected by \emph{Spitzer}, \emph{Herschel}, and the Atacama Pathfinder Experiment (APEX) was published by \citet{Furlan+2016}.
   
   Crystallized water ice in d216-0939 was detected by \citet{TeradaTokunaga2012}. Spatially resolved IR observations, probing the small grain population (up to approximately \SI{1}{\micro\meter}), reveal extended emission originating from the upper disk layers. High spectral resolution JWST observations \citep{Potapov+2025} confirm the presence of crystalline water ice. Additionally, other ices, complex organic molecules, and amorphous silicates are detected. Notably, \citet{Potapov+2025} report that scattering opacities of water ice are necessary to model the shape of the water ice absorption feature.
   
   In this study, we constrain the spatial dust distribution including the water ice abundance and crystallization state in the edge-on silhouette disk d216-0939 in the ONC, providing an estimate of its inclination angle. Additionally, we disentangle the contributions of line-of-sight (LOS) absorption of starlight, thermal reemission of the dust, and scattered radiation to the integrated spectrum. To this end, we derive constraints on the dust distribution by modeling JWST and HST observations simultaneously, performing Monte Carlo radiative transfer (MCRT) simulations. The PPD model found is then used to investigate the water ice feature observed with the \emph{Near-Infrared Spectrograph} (NIRSpec) in detail. For this purpose, we employ the spectral importance sampling method, enabling us to model the shape of the water ice feature at a high spectral resolution.

   \section{Observations and analysis}\label{sec:observations}
   JWST observations of the silhouette disk d216-0939, collected with the instruments NIRSpec and \emph{Mid-Infrared Instrument} (MIRI), were published by \citet{Potapov+2025}. They identify and fit multiple absorption features, with the most prominent being the water ice feature, at $\sim$\SI{3}{\micro\meter}, the carbon dioxide ice feature at $\sim$\SI{4.3}{\micro\meter}, and the silicates feature at $\sim$\SI{10}{\micro\meter}.

   We model the observed SED to obtain a continuum baseline, i.e., the featureless flux, to subsequently fit the water ice feature. To this end, we select the wavelengths $\lambda_\text{cont} \in \{ \num{1.67}, \num{2.15}, \num{4.0}, \num{5.2}, \num{7.89}, \num{14.85}, \num{19.99}, \num{24.07} \} \, \si{\micro\meter}$, for which we assume that there are no absorption or emission features in the spectrum. We employ the criterion
\begin{equation}
  \label{eq:best-fit_criterion}
  \chi_\text{cont}^2 = \sum_{\lambda = \lambda_\text{cont}} \log \left( \frac{F_\lambda^\text{tot;sim}}{F_\lambda^\text{tot;obs}}  \right)^2,
\end{equation}
with observed and simulated spectral flux densities, $F_\lambda^\text{tot;obs}$ and $F_\lambda^\text{tot;sim}$, respectively, to pre-select best-fit candidate models. Ultimately, the best fit is determined by comparing their simulated appearance in H$\alpha$, $\lambda_{\text{H}\alpha} = \SI{658}{\nano\meter}$, to spatially resolved HST observations. Finally, based on the best-fit model, the water ice feature is modeled at \num{141} equidistantly spaced wavelengths between \SI{2.6}{\micro\meter} and \SI{4.0}{\micro\meter}.

   \section{Disk and dust model}\label{sec:model}
   The PPD model consists of a rotationally symmetric 3D dust density distribution described in Sect.~\ref{sec:disk_structure}. The dust is modeled as a continuous grain size population of spherical compact dust grains of different chemical and mineralogical composition (see Sect.~\ref{sec:dust_properties}). As we aim to reproduce the observations from visual to mid-IR wavelengths, we only consider the small grain population in the disk. Finally, the water ice content and spatial distribution is modeled via incorporation of a snowline, as described in Sect.~\ref{sec:snowline_model}.

\subsection{Disk structure}\label{sec:disk_structure}
Viscous accretion disk theory established by \citet{ShakuraSunyaev1973} and \citet{LyndenBellPringle1974} suggests a rotational symmetric gas density prescription,
\begin{equation}
  \label{eq:lynden_bell_pringle_disk}
  \begin{aligned}
  \varrho_\text{disk}(r, z) &= \varrho_\text{ref} \left(  \frac{R_\text{ref}}{r} \right)^\alpha \exp\left[ - \frac{1}{2} \left( \frac{z}{h(r)}\right)^2 - \left( \frac{R_\text{out}}{r} \right)^{\alpha - \beta - 2} \right], \\
  h(r) &= h_\text{ref} \left( \frac{r}{R_\text{ref}}\right)^\beta,
  \end{aligned}
\end{equation}
in cylindrical coordinates $r$ and $z$\footnote{The density profile can also be written as $\varrho_\text{disk}(\Sigma(r), h(r, z))$ with $\Sigma$ denoting the vertically integrated surface density. This form is frequently found in literature \citep[e.g.,][]{Woitke+2016,BrunngraeberWolf2020,Hofmann+2022} using a surface density exponent $\gamma \equiv \alpha - \beta$.}. Eq.~\ref{eq:lynden_bell_pringle_disk} employs an exponential truncation factor consistent with the self-similar solution given in \citet{Hartmann+1998}. The parameters,
\[
\begin{array}{lp{0.8\linewidth}}
  \varrho_\text{ref} & reference density $\varrho_\text{ref} = \varrho_\text{disk}(R_\text{ref} < R_\text{out},0)$; \\
  R_\text{ref}       & reference radius; \\
  \alpha            & midplane density exponent $\varrho_\text{disk}(r<R_\text{out},0) \propto r^{-\alpha}$; \\
  R_\text{out}       & exponential density drop-off truncation radius; \\
  h_\text{ref}       & reference scale height $h_\text{ref} \equiv h(R_\text{ref})$; \\
  \beta             & flaring exponent,
\end{array}
\]
determine the density profile of the model alongside a wall-shaped inner disk rim at $R_\text{in}$ and a fixed spherical outer boundary radius $R_\text{max} = \SI{1000}{\astronomicalunit}$.

To reduce the number of free parameters it is expedient to relate the flaring and midplane density exponents via
\begin{equation}
  \alpha = 3 \big(\, \beta - \nicefrac{1}{2} \big),
\end{equation}
as suggested by \citet{LinPapaloizou1985}, and successfully used to model the appearance of PPDs \citep[e.g.,][]{Burrows+1996,Wolf+2003,Brunngraeber+2016}. Finally, the reference radius is set to $R_\text{ref} = \SI{100}{\astronomicalunit}$, the reference density $\varrho_\text{ref}$ is computed to yield a total gas mass $M_\text{disk}$ inside the simulation domain as described in Sect.~\ref{apx:mass_normalization}, and a gas-to-dust mass ratio of \num{100} is used.

\subsection{Dust properties}\label{sec:dust_properties}
Motivated by the analytical prescription for the grain size distribution of the diffuse interstellar medium \citep{Mathis+1977,WeingartnerDraine2001}, we assume that the number density, $n_\text{dust}(a)$, of grain sizes, $a$, follows the power law $dn_\text{dust} (a) \propto a^{\num{-3.5}} \, da$, and include grain sizes of $a_\text{min} \leq a \leq a_\text{max}$ with $a_\text{min} = \SI{5}{\nano\meter}$ fixed. We account for grain growth in the PPD \citep[e.g.,][]{DullemondDominik2005,Natta+2007,Dartois+2022} by considering $a_\text{max} \geq \SI{250}{\nano\meter}$.

To derive optical properties, i.e., mass opacities, $\varkappa_\lambda^\text{ext} = \varkappa_\lambda^\text{sca} + \varkappa_\lambda^\text{abs}$, and scattering matrices $\mathbf{F}_\lambda(\psi)$, of an effective medium dust grain population from measured refractory indices, we combine sets of $(n, \kappa)$-values per dust component into a single set of refractory indices using the Bruggeman mixing rule \citep[e.g.,][]{BohrenHuffman1998}. Successively, we compute optical properties using the \texttt{miex} algorithm \citep{WolfVoshchinnikov2004}, which employs Mie theory. Finally, we use grain size-averaged optical parameters as described in \citet{Wolf2003}.

We use the DSHARP \citep{Birnstiel+2018} dust components without water ice, i.e., refractory organics \citep[$\rho_c = \SI{1.5}{\gram\per\centi\meter\cubed}$;][]{HenningStognienko1996}, astronomical silicates \citep[``Astrosil''; $\rho_c = \SI{3.3}{\gram\per\centi\meter\cubed}$;][]{Draine2003b}, and troilite \citep[$\rho_c = \SI{4.83}{\gram\per\centi\meter\cubed}$;][]{HenningStognienko1996}, to model the chemical composition of the dust. The mass fractions of the DSHARP (no ice) dust model constituents are given in Table~\ref{tab:ice_mixtures} and their optical properties are shown in Fig.~\ref{fig:dsharp_decomposition}.

To assess the effect of irregularly shaped grains compared to the spherical compact grain shape intrinsic to the Mie approach, we additionally calculated optical properties using a distribution of hollow spheres \citep{Min+2005} with \texttt{optool} \citep{Dominik+2021} assuming a maximum vacuum volume fraction of \num{0.8}. While the scattering opacities are moderately offset to lower values and the shape of the silicates feature at $\sim$\SI{10}{\micro\meter} is slightly broadened, the general shapes of features in the DSHARP (no ice) and the icy dust mixtures described in Sect.~\ref{sec:snowline_model} remain unchanged.

\begin{table*}[h]
  \centering
  \caption[]{\label{tab:ice_mixtures}Effective medium optical constants of different (icy) dust mixtures used in this work.}
  \small
 \begin{tabular}{lccccc}
   \toprule
   Dust mixture & \multicolumn{4}{c}{Mass fractions $f_c$} & Bulk density \\ \addlinespace[.5em]
   
   \# & Astrosil & MgSiO\textsubscript{3}/H\textsubscript{2}O & Organics & Troilite & $\rho_\text{mix}$ \\
   \midrule
   0: DSHARP (no ice) & \num{0.411} & -- & \num{0.496} & \num{0.093} & \SI{2.11}{\gram\per\centi\meter\cubed} \\
   \midrule
   1: \SI{11}{\percent} ice; \SI{100}{\kelvin} & -- & \num{0.300}/\num{0.111} & \num{0.496} & \num{0.093} & \SI{1.70}{\gram\per\centi\meter\cubed} \\
   2: \SI{11}{\percent} ice; \SI{150}{\kelvin} & -- & \num{0.300}/\num{0.111} & \num{0.496} & \num{0.093} & \SI{1.70}{\gram\per\centi\meter\cubed} \\
    \midrule
   3: \ \ \SI{6}{\percent} ice; \SI{100}{\kelvin} & \num{0.206} & \num{0.150}/\num{0.056} & \num{0.496} & \num{0.093} & \SI{1.88}{\gram\per\centi\meter\cubed} \\
   4: \ \ \SI{6}{\percent} ice; \SI{150}{\kelvin} & \num{0.206} & \num{0.150}/\num{0.056} & \num{0.496} & \num{0.093} & \SI{1.88}{\gram\per\centi\meter\cubed} \\
   \midrule
   5: \ \ \SI{3}{\percent} ice; \SI{100}{\kelvin} & \num{0.309} & \num{0.075}/\num{0.028} & \num{0.496} & \num{0.093} & \SI{1.99}{\gram\per\centi\meter\cubed} \\
   6: \ \ \SI{3}{\percent} ice; \SI{150}{\kelvin} & \num{0.309} & \num{0.075}/\num{0.028} & \num{0.496} & \num{0.093} & \SI{1.99}{\gram\per\centi\meter\cubed} \\
  \bottomrule
 \end{tabular}
 \tablefoot{The icy mixtures are created by replacing the total, half, and quarter amount of Astrosil in the DSHARP (no ice) dust mixture by the MgSiO\textsubscript{3}/H\textsubscript{2}O component, resulting in water-ice contents of \SI{11}{\percent}, \SI{6}{\percent}, and \SI{3}{\percent} by mass, respectively.}
\end{table*}

\subsection{Free parameter space}\label{sec:parameter_space}
We chose the value ranges of the free parameters, listed in Table~\ref{tab:parameter_space}, based on the following assumptions: i) The luminosity, $L_\star$, of the host star is in the order of magnitude found by \citet{Factor+2017} and \citet{Sheehan+2020}. ii) The JWST observations in the near- and mid-IR probe the small grain population in the upper and outer disk layers. Therefore, due to dust settling \citep[e.g.,][]{Dubrulle+1995,SchraeplerHenning2004,BrunngraeberWolf2020,Sturm+2023a,Dartois+2025}, the reference scale height, $h_\text{ref}$, is higher, and the disk mass, $M_\text{disk}$, is lower, compared to the corresponding parameter values of the large grain population inferred by \citet{Sheehan+2020} from ALMA observations. iii) The maximum temperature at the inner rim of the disk is the presumed dust sublimation temperature, $T_\text{sub}$, of $\sim$\SI{1500}{\kelvin} \citep[e.g.,][]{Vaidya+2009,Menu+2015,Hofmann+2022} yielding an inner rim radius of \SI{0.05}{\astronomicalunit}. The parameter space is sampled for $N_\text{values}$ distinct values of the free parameters given in Table~\ref{tab:parameter_space}.

\begin{table*}
\caption{Chosen value ranges for the exploration of the parameter space alongside published values found in literature.}             
\label{tab:parameter_space}      
\centering
\sisetup{
  range-phrase = {--},
  separate-uncertainty = true,
  range-units = single
}
\begin{tabular}{cccccc}     
  \toprule       
  Parameter & Value ranges & $N_\text{values}$ & Fixed & Literature values & References \\ 
  \midrule
  Distance $d_\star$ & -- & -- & \cmark & \SI{362}{pc} & 1  \\
  Effective temperature $T_\star$ & -- & -- & \cmark & $\SI{4140}{\kelvin}$ (spectral type K5) & 2,3  \\ \addlinespace[.5em]
  \multirow{3}{*}{Luminosity $L_\star$} & \multirow{3}{*}{\SIrange{.75}{1.25}{\Lsun}} & \multirow{3}{*}{\num{3}} & \multirow{3}{*}{\xmark} & \SI{0.153}{\Lsun} & 3 \\
  & & & & \SI{0.97}{\Lsun} & 4 \\
  & & & & \SI{0.57}{\Lsun} & 5 \\
  \midrule
  Inclination $\iota$ & \SIrange{72.5}{77.5}{\degree} & \num{3} & \xmark & \SI{77.7}{\degree} & 4 \\
  Position angle $\PA$ & -- & -- & \cmark & \SI{83}{\degree} & 6 \\
  \midrule
  Minimum grain size $a_\text{min}$  & -- & -- & \cmark & \SI{5}{\nano\meter} & 7,8 \\
  Maximum grain size $a_\text{max}$  & \SIrange{0.25}{5}{\micro\meter} & \num{3} & \xmark & \SI{250}{\nano\meter} & 7,8 \\
  Grain size exponent $a_\text{pow}$ & -- & -- & \cmark & \num{3.5} & 7,8 \\
  \midrule
  Inner radius $R_\text{in}$ & \SI{.05}{\astronomicalunit} & -- & \cmark & $T_\text{sub} \approx \SI{1500}{\kelvin}$ & 9,10 \\ \addlinespace[.5em]
  \multirow{3}{*}{Outer radius $R_\text{out}$} & \multirow{3}{*}{\SIrange{200}{500}{\astronomicalunit}} & \multirow{3}{*}{\num{4}} & \multirow{3}{*}{\xmark} & \SI{291}{\astronomicalunit} & 11 \\
  & & & & \SI{572}{\astronomicalunit} & 12 \\
  & & & & \SI{600}{\astronomicalunit} & 6 \\ \addlinespace[.5em]
  \multirow{3}{*}{Disk mass $M_\text{disk}$} & \multirow{3}{*}{\SIrange[print-unity-mantissa=false]{1e-4}{2e-3}{\Msun}} & \multirow{3}{*}{\num{4}} & \multirow{3}{*}{\xmark} & \SI{1.93e-3}{\Msun} & 11  \\
  & & & & {\color{gray}\SI{4.5e-2}{\Msun}} & 4 \\
  & & & & {\color{gray}\SI{0.399}{\Msun}} & 12 \\ \addlinespace[.5em]
  Scale heigh $h_\text{ref}(R_\text{ref} = \SI{100}{\astronomicalunit})$ & \SIrange{12.5}{17.5}{\astronomicalunit} & \num{3} & \xmark & {\color{gray}\SI{10.3}{\astronomicalunit}} & 4 \\
  Flaring exponent $\beta$ & \numrange{1.125}{1.25} & \num{2} & \xmark & -- & 13,14 \\
\bottomrule                  
\end{tabular}
\tablefoot{Parameter values in the given value ranges are chosen equidistantly on a linear scale, with $N_\text{values}$ steps, except the disk mass, which is spaced logarithmically, and the maximum grain size, for which we use \SI{0.25}{\micro\meter}, \SI{1}{\micro\meter}, and \SI{5}{\micro\meter}. The literature values grayed out were derived from observations probing the large grain population in the disk and are therefore not considered in the chosen value ranges.}
\tablebib{(1)~\citet{GaiaEDR3};
  (2) \citet{Hillenbrand1997}; (3) \citet{PecautMamajek2013}; (4) \citet{Sheehan+2020};
  (5) \citet{Furlan+2016}; (6) \citet{Smith+2005}; (7) \citet{Mathis+1977};
  (8) \citet{WeingartnerDraine2001}; (9) \citet{Vaidya+2009}; (10) \citet{Hofmann+2022};
  (11) \citet{MannWilliams2009}; (12) \citet{TeradaTokunaga2012}; (13) \citet{KenyonHartmann1987};
  (14) \citet{Wolf+2003}.
}
\end{table*}

\subsection{Water ice inclusion}\label{sec:snowline_model}
While icy grains are often modeled as dust cores covered by an ice mantle \citep[e.g.,][]{Dartois2006,Dartois+2022,Sturm+2023b}, we use measured $(n, \kappa)$-values of silicate-water ice mixtures \citep[MgSiO\textsubscript{3}/H\textsubscript{2}O;][see Sect.~\ref{apx:effective_medium_icy_dust}]{Potapov+2018,Potapov+2021} to include water ice in the dust model, which is motivated by the best-fit chemical composition of the observed water ice feature found by \citet{Potapov+2025}. Consequently, we use MgSiO\textsubscript{3}/H\textsubscript{2}O optical constants, $(n, \kappa)$, with a silicate-to-water ice mass ratio of \num{2.7} \citep{Potapov+2018}. We consider water ice in amorphous and crystalline form using \SI{100}{\kelvin} and \SI{150}{\kelvin} silicate-water ice optical constants, $(n, \kappa)$, measured at the respective temperatures, as the onset of crystallization occurs between \SI{120}{\kelvin} and \SI{140}{\kelvin} \citep[e.g.,][]{JenniskensBlake1994,Mifsud+2022}.

\begin{figure}
  \centering
  \includegraphics[width=\columnwidth]{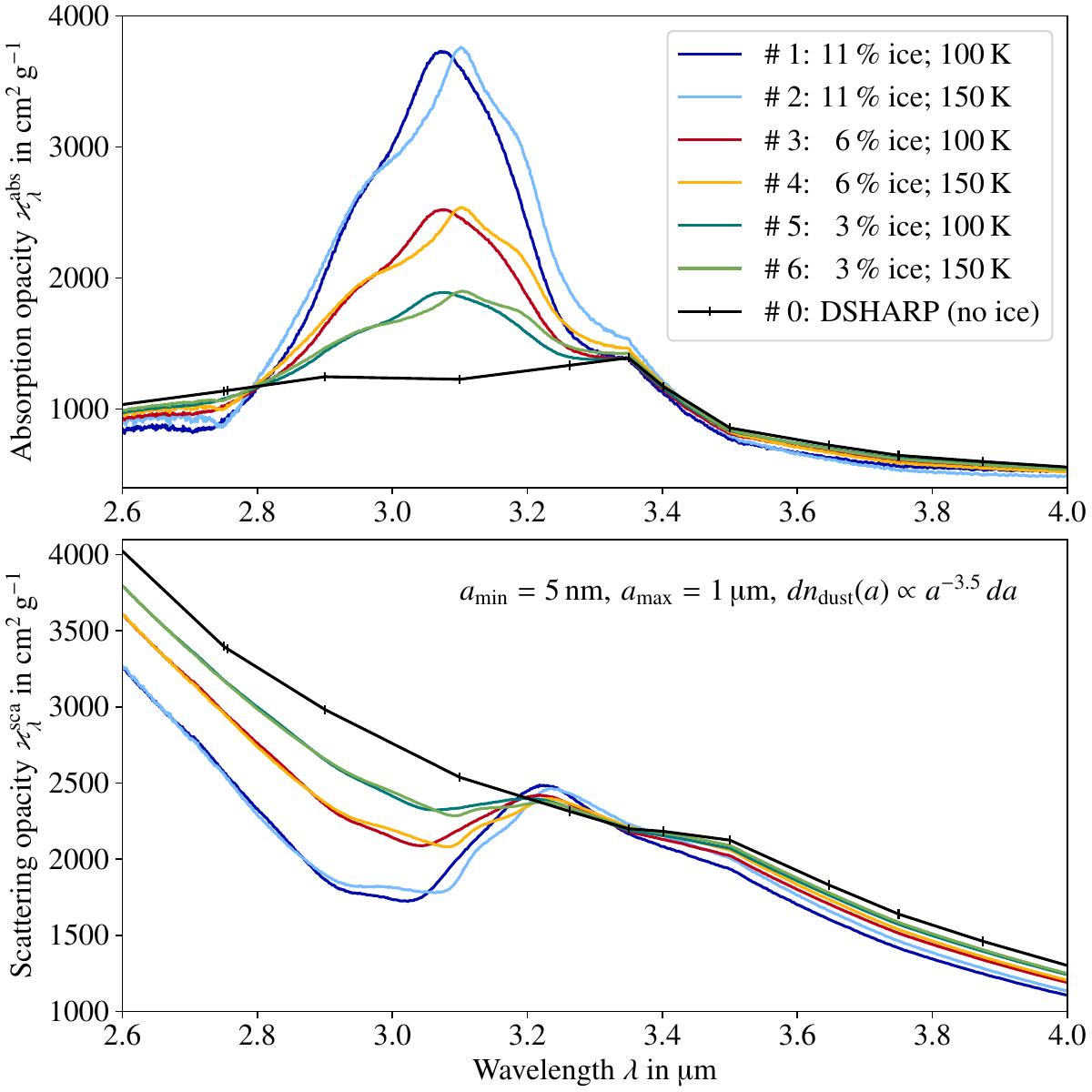}
  \caption{Effective medium opacities for the (icy) dust mixtures used in this work.}
  \label{fig:ice_mix_opacities}
\end{figure}

Following a pragmatic approach, we create effective medium dust mixtures with varying water ice content by (partially) replacing the Astrosil component of the DSHARP (no ice) dust mixture by the MgSiO\textsubscript{3}/H\textsubscript{2}O component. The resulting dust mixtures used in this work are listed in Table~\ref{tab:ice_mixtures}, and the corresponding opacities are shown in Fig.~\ref{fig:ice_mix_opacities}.

The spatial distribution of water ice in the disk model (Sect.~\ref{sec:disk_structure}) is assumed to be temperature dependent. As the optical constants of the MgSiO\textsubscript{3}/H\textsubscript{2}O material are only provided in the near- and mid-IR, i.e., wavelengths of \SIrange{1.7}{25}{\micro\meter}, which is insufficient to compute the thermal interaction of dust with stellar radiation, we compute the dust temperature distribution using the DSHARP (no ice) dust mixture. Assuming that the dust temperature is not significantly influenced by the presence of water ice, we place water ice in the disk in regions beyond the snowline, for which we assume $T_\text{snow} = \SI{170}{\kelvin}$ \citep[e.g.,][]{MartinLivio2012}.

   \section{Radiative transfer simulations}
   Our modeling approach is based on RT simulations to calculate the SED and spatially resolved images for the disk model within the considered parameter space. Both scattered radiation as well as thermal reemission radiation of the dust, based on self-consistently calculated spatial dust temperature distributions, are considered.
   
   \subsection{Computation of simulated fluxes}\label{sec:methods_comp_sim_fluxes}
   We use the MCRT code \texttt{POLARIS} \citep{Reissl+2016,Reissl+2018} to compute synthetic images at single wavelengths. The computation of the thermal equilibrium \citep[e.g.,][]{Wolf+1999} of the dust irradiated by a given radiation source, i.e., a star modeled as a blackbody point source, is computed via the \cite{BjorkmanWood2001} algorithm. Successively, we split the flux,
  \begin{equation}
    \label{eq:flux_splitting}
    F_\lambda^\text{tot} = F_\lambda^{\text{emi,}\star} + F_\lambda^\text{emi,dust} + F_\lambda^{\text{sca,}\star} + F_\lambda^\text{sca,dust},
  \end{equation}
  and compute each of the contributions separately taking the corresponding source term into account (see Sect.~\ref{apx:radiative_transfer}). The contributions to the total flux, $F_\lambda^\text{tot}$, are the emission of the host star attenuated by dust in the LOS, $F_\lambda^{\text{emi,}\star}$, the thermal emission of the dust attenuated by dust in the LOS to the detector pixels, $F_\lambda^\text{emi,dust}$, the scattered starlight, $F_\lambda^{\text{sca,}\star}$, and the scattered dust emission, $F_\lambda^\text{sca,dust}$. We employ a ray-tracing algorithm to compute the flux contributions $F_\lambda^{\text{emi,}\star}$ and $F_\lambda^\text{emi,dust}$ and sample the scattering source terms, $S_\lambda^{\text{sca,}\star}$ and $S_\lambda^\text{sca,dust}$, via Monte Carlo simulations utilizing enforced first scattering \citep{Mattila1970} and peel-off \citep{YusefZadeh1984} to obtain the scattered flux contributions, $F_\lambda^{\text{sca,}\star}$ and $F_\lambda^\text{sca,dust}$.

  \subsection{Spectral importance sampling}\label{sec:spectral_importance_sampling}
  The computational cost to sample the scattering source terms is high, which limits the achievable spectral resolution as the solution of the RT equation is computed at each wavelength, $\lambda$, independently. To overcome these limitations, we employ spectral importance sampling.
  
  The concept of spectral importance sampling is to estimate the fluxes in a wavelength bin of width $\Delta \lambda$ around a wavelength $\lambda_\text{MC}$, i.e., $2 |\lambda - \lambda_\text{MC}| \leq \Delta \lambda$, by applying an appropriate correction to the photon package (PP) weights at wavelengths $\lambda \neq \lambda_\text{MC}$ \citep[e.g.,][]{Jonsson2006,Emde+2011}. As the propagation of PPs is the computationally most expensive part of MCRT simulations, utilizing polychromatic PPs allows to increase the spectral resolution significantly. While the implementation, validation, and performance of the spectral importance sampling method are described in Sect.~\ref{apx:gespenst}, its concept is briefly summarized in the following.

   \subsubsection{Scattering simulations}
   In the extinction framework \citep[e.g.,][]{KriegerWolf2023}, the update of a PP weight during a scattering event, i.e., $I_{\lambda,i}^\text{PP} \rightarrow I_{\lambda,i+1}^\text{PP}$, at a position $\vec{r}$ after traveling the path from $s_0$ to $s$ on a straight ray through a dusty medium is descibed by 
   \begin{equation}
     \label{eq:pp_weight_update_path}
     I_{\lambda,i+1}^\text{PP} = \underbrace{ \frac{\varkappa_\lambda^\text{sca}(\vec{r})}{\varkappa_\lambda^\text{ext}(\vec{r})} }_{\text{scattering albedo}} \underbrace{ \frac{\varkappa_\lambda^\text{ext}(\vec{r}) e^{-\tau_{\lambda}(s)} }{\varkappa_{\lambda_\text{MC}}^\text{ext}(\vec{r}) e^{-\tau_{\lambda_{\text{MC}}}(s)} } }_{\text{weight correction}}  I_{\lambda,i}^\text{PP},
   \end{equation}
   with the optical depths (Eq.~\ref{eq:optical_depth}), $\tau_{\lambda_{\text{MC}}}(s)$ and $\tau_{\lambda}(s)$, numerically evaluated. This correction accounts for the ``wrong'' position and traveled optical depth of the PP at wavelength $\lambda$ compared to the wavelength $\lambda_\text{MC}$. For anisotropic scattering (e.g., Mie scattering), an additional weight correction is necessary to take the difference in probability to scatter into a certain direction, expressed via the scattering angles $\psi$ and $\xi$, into account:
   \begin{equation}
     \label{eq:pp_weight_update_anisotropy}
     I_{\lambda,i+1}^\text{PP,Mie} = \underbrace{\frac{p_\lambda(\psi) p_\lambda(\xi)}{p_{\lambda_\text{MC}}(\psi)p_{\lambda_\text{MC}}(\xi)}}_{\text{anisotropy weight correction}} I_{\lambda,i+1}^\text{PP},
   \end{equation}
   with the probability density functions $p_\lambda(\psi)$, $p_\lambda(\xi)$, $p_{\lambda_\text{MC}}(\psi)$, and $p_{\lambda_\text{MC}}(\xi)$ denoting the probabilities at certain wavelengths to scatter into a certain angle.

   \subsubsection{Probabilistic dust emission}
   Considering the heated dust as a radiation source to sample the source terms $S_\lambda^\text{emi,dust}$ and $S_\lambda^\text{sca,dust}$ (see Sect.~\ref{apx:radiative_transfer}), PPs are emitted according to a spatial distribution following the local monochromatic luminosity,
   \begin{equation}
     L_\lambda(V) = 4 \pi \varrho_\text{dust} V \varkappa_\lambda^\text{abs} B_\lambda(T),
   \end{equation}
   of dust in a volume $V$ with density $\varrho_\text{dust}$, mass absorption opacity $\varkappa_\lambda^\text{abs}$, and the Planck function $B_\lambda(T)$ of the dust with temperature $T$ \citep[e.g.,][]{Wolf+1999}. Employing the spectral importance sampling method, the initial intensity of a PP,
   \begin{equation}
     I_{\lambda,0}^{\text{PP}} = \frac{L_\lambda(V)}{L_{\lambda_\text{MC}}(V)} I_{\lambda_\text{MC},0}^\text{PP},  
   \end{equation}
   is corrected to account for the difference in probability of a PP to be emitted from a certain volume compared to the wavelength $\lambda_\text{MC}$.

\section{Results}\label{sec:results}
Based on the reference model described in the following, we constrain the spatial distribution, ice content, and crystallization state of water ice in the disk via a fit to the water ice feature.

\subsection{Reference model}\label{sec:continuum_baseline_disk_model}

\begin{figure}
  \centering
  \includegraphics[width=\columnwidth]{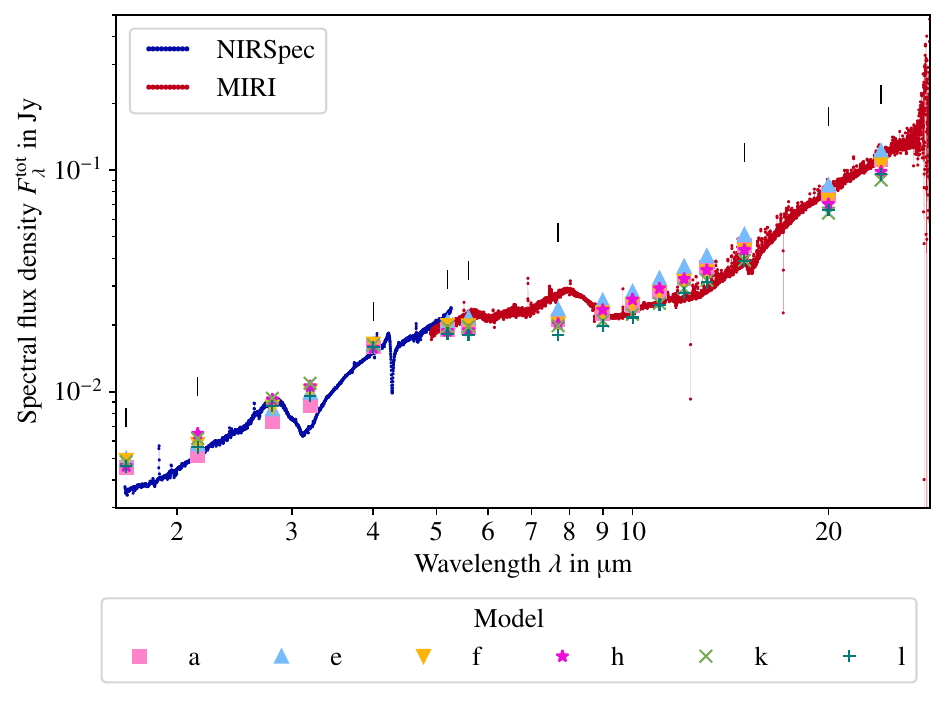}
  \caption{Comparison of the observed SED to the well-fitting disk models considered as best-fit candidates. The free parameter values of the models are listed in Table \ref{tab:best-fit_continuum_models}. The wavelengths $\lambda_\text{cont}$ are marked by vertical black lines above the data.}
  \label{fig:best-fit_candidates}
\end{figure}

The parameter space (Table~\ref{tab:parameter_space}) is highly degenerate with respect to the fit to the SED. While the maximum grain size, flaring exponent, stellar luminosity, and reference scale height are constrained to $a_\text{max} = \SI{1}{\micro\meter}$, $\beta=\num{1.250}$, $L_\star > \SI{0.75}{\Lsun}$, and $h_\text{ref} > \SI{12.5}{\astronomicalunit}$, the disk mass, outer radius, and inclination remain unconstrained within the explored parameter ranges. We assessed the suitability of 12 best-fit candidates with $\chi^2_\text{cont} < 0.378$, labeled a, b, c, d, e, f, g, h, k, and l, by increasing $\chi^2_\text{cont}$, to represent a continuum baseline in the wavelength range \SIrange{2.6}{4.0}{\micro\meter}, selecting models a, e, f, h, k, and l to compare to HST observations. The selected and rejected models are shown in Figs.~\ref{fig:best-fit_candidates} and \ref{fig:best-fit_candidates_deselected}, respectively, with their parameter values listed in Table~\ref{tab:best-fit_continuum_models}.

To produce synthetic observations comparable to the HST H$\alpha$-image \citep{Smith+2005} a constant background flux intensity $I_\text{bg}(\lambda) = f_\text{bg} \times B_\lambda(T_\text{bg})$, with a scaling factor, $f_\text{bg}$, and the blackbody function $B_\lambda$, is incorporated into the radiative transfer modeling, neglecting scattering of background radiation. We assume a background temperature of $T_\text{bg} = \SI{10000}{\kelvin}$, as expected in the outskirts of the Orion Nebula \citep{Balick+1974}.

\begin{figure*}
  \centering
  \includegraphics[width=\textwidth]{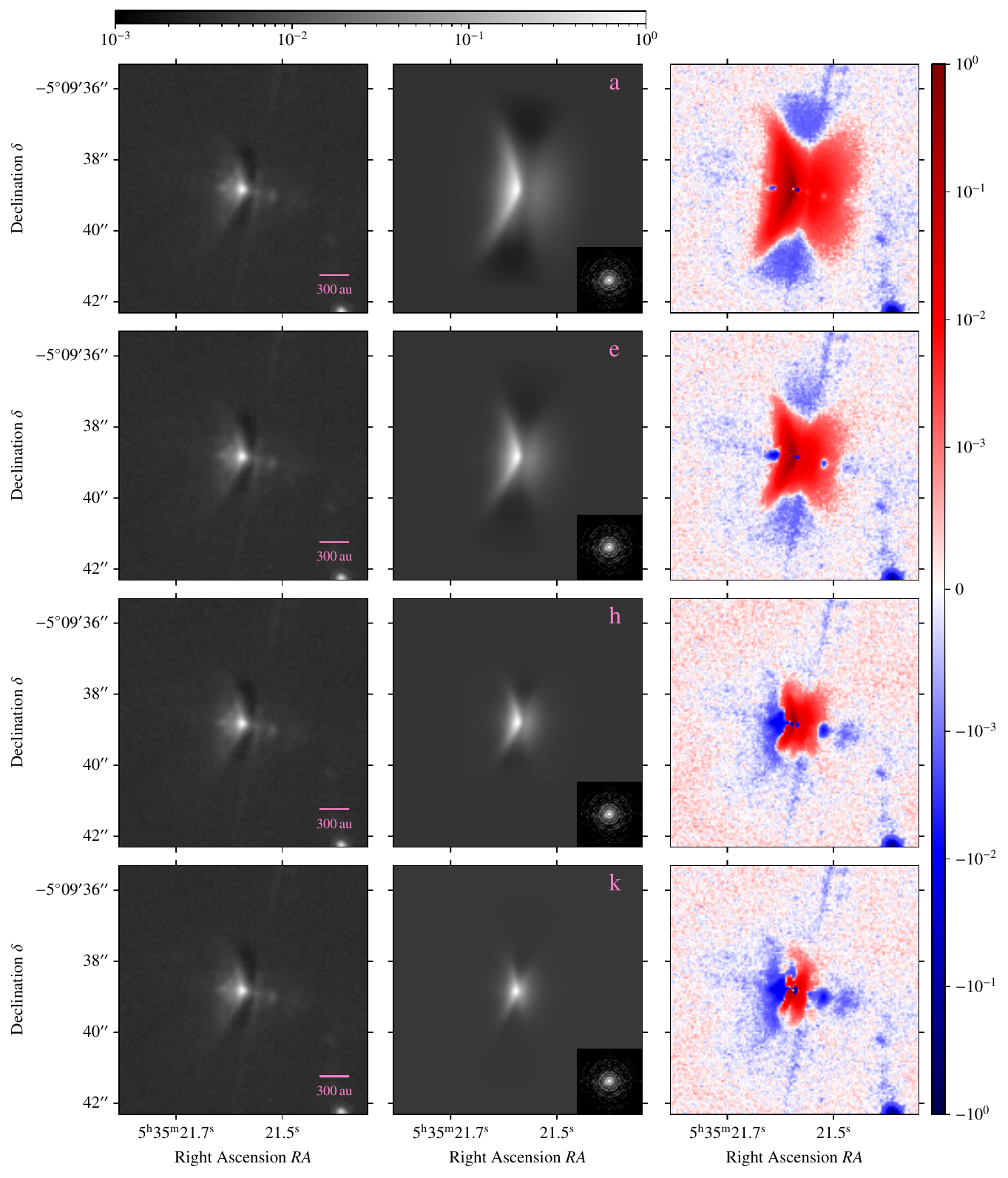}
  \caption{Comparison of HST observations at $\lambda_{\text{H}\alpha} = \SI{658}{\nano\meter}$ with synthetic images for selected disk models (see Table \ref{tab:best-fit_continuum_models}). After convolution, the images in the middle panel were downsampled to the resolution of the images in the left column with a flux conserving algorithm \citep{DeForest2004}. All grayscale images are normalized with respect to their maximum. \emph{Left column:} HST H$\alpha$-image by \citet{Smith+2005} with a scalebar indicating the image dimension for a distance of $d_\star = \SI{362}{\parsec}$. \emph{Middle column:} Synthetic images produced by convolution of simulation results with the appropriate F658N filter point spread function of HST \citep[created using \texttt{TinyTim};][]{Krist+2011}, which is depicted in the lower right corner of each image on a normalized logarithmic grayscale ranging from $\num[print-unity-mantissa=false]{1e-4}$ to $\num{1}$. \emph{Right column:} Difference images obtained by subtraction of the simulation results from the observations. Red indicates a flux overestimation by our modeling and blue the opposite.}
  \label{fig:hst_compare}
\end{figure*}

The appearance of the disk in H$\alpha$ for pre-selected models is shown next to the HST observations in Fig.~\ref{fig:hst_compare}. Background intensity scaling factors of \num{1e-13}, \num{1.75e-13}, \num{2.25e-13}, and \num{4e-13} were used for the models a, e, h, and k, respectively.

The comparison of normalized fluxes\footnote{As we do not consider line emission, the absolute flux values cannot be compared.} shows that the models a and e overestimate the spatial extent of the emission significantly (which is also the case for the models f and l). The outer radius is therefore constrained to $R_\text{out} \leq \SI{400}{\astronomicalunit}$. Based on visual inspection of the shape and extent of the emission region, we select model k with an inclination angle of \SI{77.5}{\degree} as best fit to the continuum. The synthetic image of model h ($\iota = \SI{72.5}{\degree}$) is also in good agreement with the HST observations. However, the flux overestimation extends over a larger area and reaches higher amplitudes above the disk midplane than for model k, as shown in the difference images in Fig.~\ref{fig:hst_compare}. Although we therefore consider model k the best fit, the results presented in Sect.~\ref{sec:water_feature_modeling} were tested for robustness by using model h as the reference model (see Sect.~\ref{sec:water_ice_feature_fit}).

The best-fit candidate models, shown in Fig.~\ref{fig:best-fit_candidates}, systematically overestimate the flux density at shorter wavelengths. Considering interstellar extinction, i.e., assuming a visual extinction of $A_V = \SI{0.72}{\magnitude}$ \citep{Hillenbrand1997}, a reddening constant $R_V = 3.1$ \citep{SchultzWiemer1975}, and the extinction law by \citet{Cardelli+1989}, the flux is attenuated by \SI{11}{\percent} at $\lambda = \SI{1.67}{\micro\meter}$, and by less than \SI{5}{\percent} for wavelengths of $\lambda \geq$ \SI{2.8}{\micro\meter}, improving the fit of modeled SEDs to the observations.

Additionally, we underestimate the flux at wavelengths from \SIrange{7}{8}{\micro\meter}, effectively underestimating the depth of the observed silicate absorption feature at wavelengths of $\sim$\SI{10}{\micro\meter}.

The outer radius of the best-fit model, $R_\text{out}=\SI{400}{\astronomicalunit}$, is smaller than the values estimated by \citet{Smith+2005} and \citet[][see Table \ref{tab:parameter_space}]{TeradaTokunaga2012}. The difference in numerical values can be partially attributed to the different definitions of the outer radius, i.e., the exponential density drop-off radius versus the apparent outer edge of the silhouette disk. However, as shown in Fig.~\ref{fig:hst_compare}, model~k underestimates the spatial extent of the disk shadow against the background emission, and overestimates flux scattered on the far-side surface of the disk. A denser disk midplane, that would block background radiation and light originating from the backside of the disk, could explain the mismatch between model and observations. As the JWST spectra probe the upper and outer disk layers, which are well-constrained by our fitting approach, a more complex model including a denser midplane constrained by spatially resolved observations is not necessary for the following analysis of the water ice feature.

The outer radius of the distribution of the large grain population found by \citet{MannWilliams2009} is smaller than the outer radius of the distribution of the small grain population, which is expected, as larger grains drift inward \citep[e.g.,][]{Weidenschilling1977,Birnstiel2024}.

Finally, the lack of extended emission above the disk in the synthetic intensity map of model~k compared to the HST observations could be attributed to remnants of an infalling envelope \citep[e.g.,][]{Ulrich1976,Lietzow-Sinjen+2025,Martin+2026} or ejections feeding the known jet HH 667 \citep{Smith+2005}.

\subsection{Impact of water ice in the disk on the SED}\label{sec:water_feature_modeling}

Based on the dust temperatures of the reference model, we include water ice into the model beyond the snowline (see Sect.~\ref{sec:snowline_model}). Successively, scattering and dust emission simulations are performed to compute flux densities employing the spectral importance sampling method with a wavelength binwidth of $\Delta \lambda = \SI{0.1}{\micro\meter}$ (see Sects.~\ref{sec:spectral_importance_sampling} and \ref{apx:gespenst}).

\subsubsection{Disk temperature structure}
\begin{figure}
  \centering
  \includegraphics[width=\columnwidth]{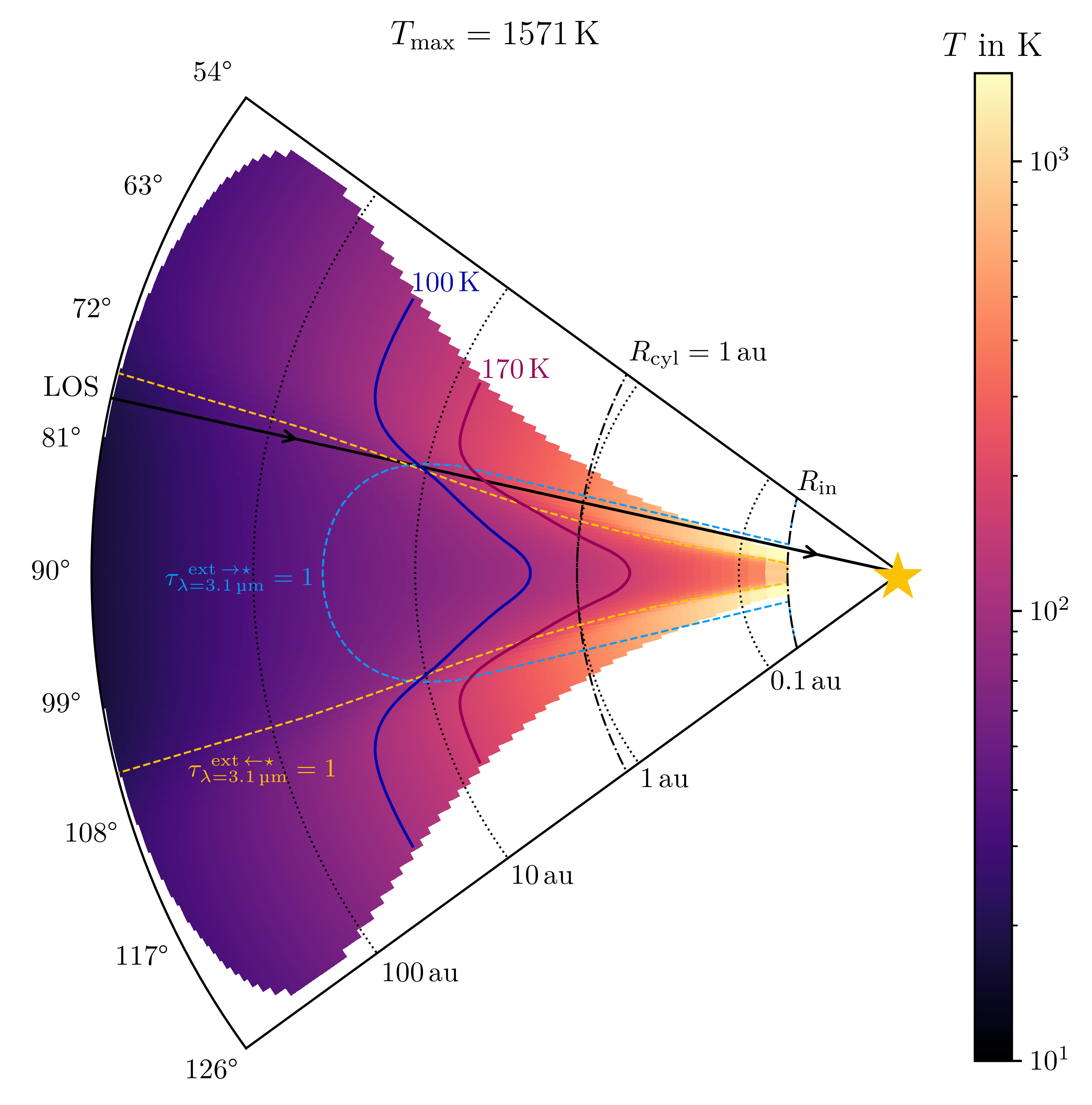}
  \caption{Vertical cut through the best-fit disk model on a logarithmic radial scale. Shown are temperatures, where the gas density, $\varrho_\text{disk}$, exceeds an assumed ISM density of $\varrho_\text{ISM} = \SI{2.7e-24}{\gram\per\centi\meter\cubed}$ \citep{Ferriere2001} by a factor of \num{1000}. The snowline at $T_\text{snow} = \SI{170}{\kelvin}$ and the \SI{100}{\kelvin}-line, as well as the unity optical depth lines at a wavelength of \SI{3.1}{\micro\meter} (computed for the best-fit model described in Sect.~\ref{sec:water_ice_feature_fit}) measured from the star to the observer and from observer to the star, illustrate the temperature and dust density structure of the disk.}
  \label{fig:temperature_structure}
\end{figure}

The dust temperatures, $T$, of the best-fit model, k, are shown in Fig.~\ref{fig:temperature_structure}. The position of the snowline at $T_\text{snow} = \SI{170}{\kelvin}$ strongly depends on the elevation above the midplane, as the inner rim of the disk shields the disk regions close to the midplane from irradiation by the host star. Water ice condenses at $\sim$\SI{5}{\astronomicalunit}, measured from the central star along the LOS, and dust temperatures drop rapidly to \SI{100}{\kelvin} at $\sim$\SI{10}{\astronomicalunit}. In the midplane of the disk, water ice can condense in-place at distances of $\lesssim$\SI{1}{\astronomicalunit} from the central star. Considering, that water ice anneals into a crystalline phase at $\sim$\SI{130}{\kelvin} \citep{JenniskensBlake1994,Mifsud+2022}, the region in the disk model, in which we assume crystalline water ice to form, is small compared to the disk dimensions.

\subsubsection{Dust model fit to the water ice feature}\label{sec:water_ice_feature_fit}
\begin{figure}
  \centering
  \includegraphics[width=\columnwidth]{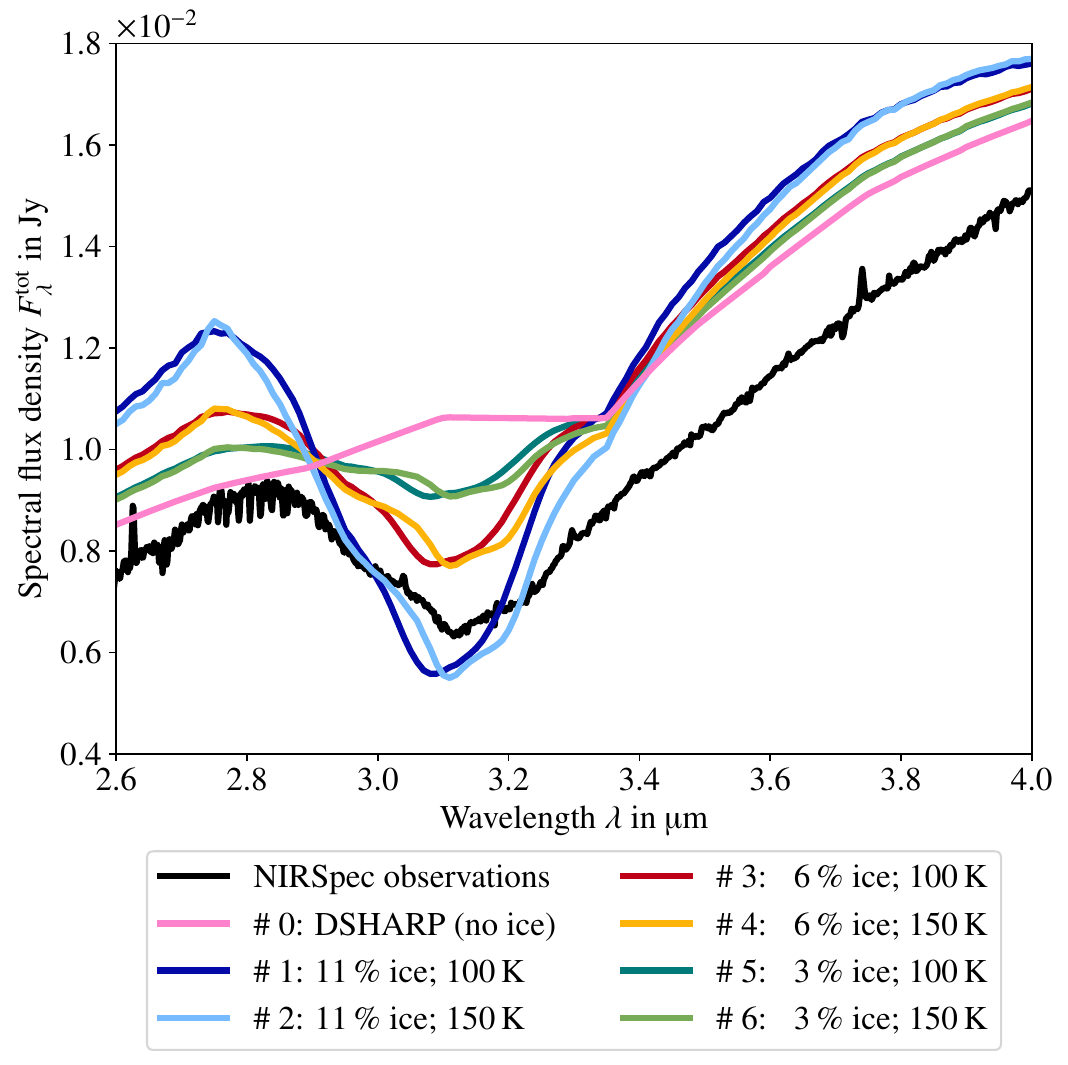}
  \caption{SED of the reference model (k) with DSHARP (no ice) dust replaced by the dust mixtures \# 1--6 beyond the snowline.}
  \label{fig:water_feature_F_tot_only}
\end{figure}

To constrain the abundance and crystallinity of water ice we performed MCRT simulations using the dust mixtures \# 1--6 listed in Table \ref{tab:ice_mixtures}. The resulting SEDs are shown in Fig.~\ref{fig:water_feature_F_tot_only}. The flux overestimation, with respect to the reference model, at wavelengths outside the water ice feature can be attributed to the difference in scattering opacities (see Fig.~\ref{fig:ice_mix_opacities}), which impacts the LOS extinction. The position of the flux minimum matches the observations only for the \SI{150}{\kelvin} silicate-water ice mixtures, i.e., the dust mixtures containing crystalline ice. The synthetic SED of dust mixture \# 4 resembles the observed SED best with respect to the shape and depth of the water ice feature. 

\begin{figure}
  \centering
  \includegraphics[width=\columnwidth]{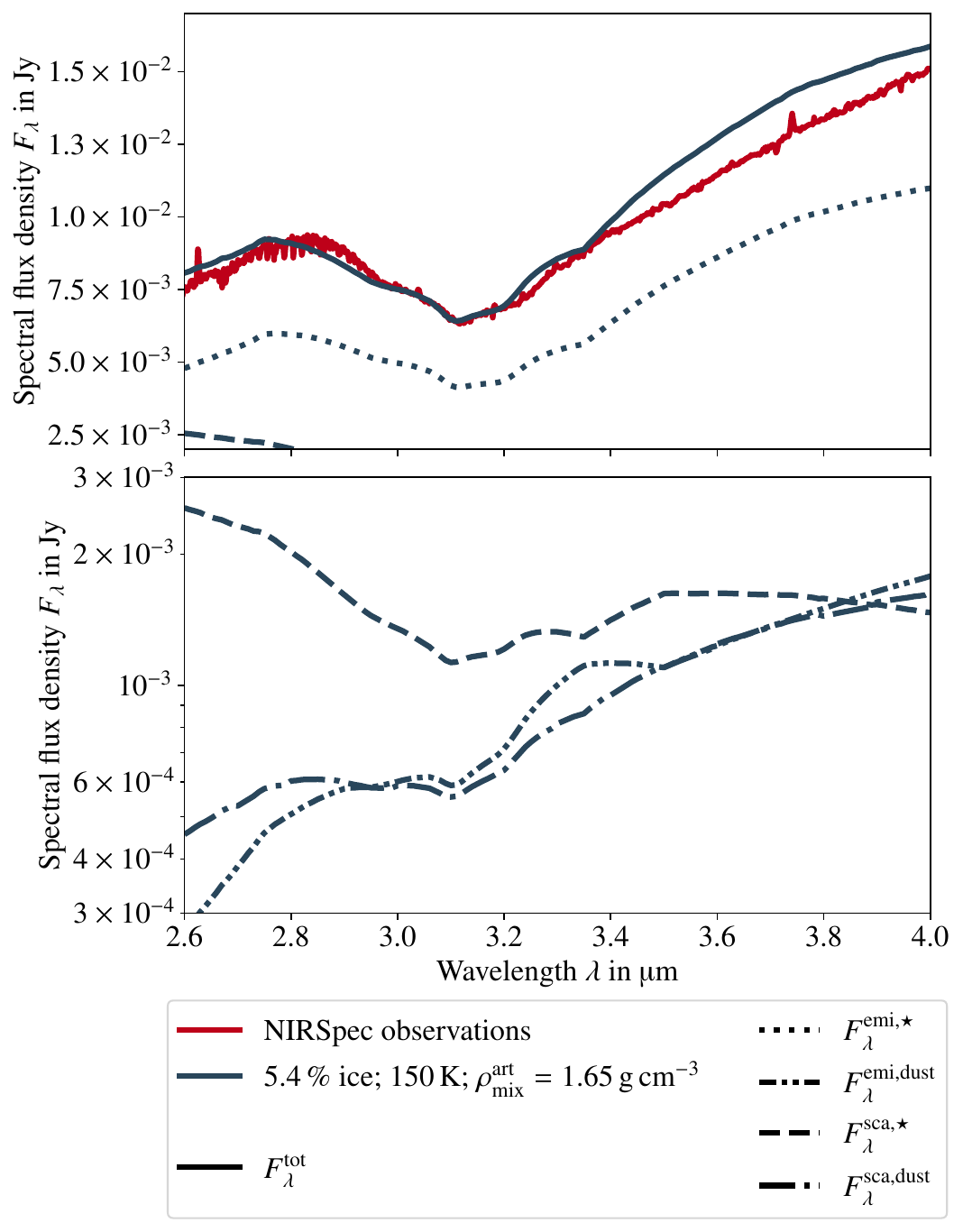}
  \caption{Best fit of the water ice absorption feature and flux contributions by source term. The black lines in the legend indicate the linestyle of the corresponding flux contributions, which are shown for the best-fit model in the upper and lower panels.}
  \label{fig:water_feature_best_fit}
\end{figure}

To improve the fit to the observations, we artificially lower the bulk density of dust mixture \# 4 and adjust its water-ice content. The artificially lowered density is effectively a dust density discontinuity at the snowline, which is physically justified as water is expected to freeze out from the gas phase and adds up to the total dust mass. We found that an artificial dust density of $\rho_\text{mix}^\text{art} = \SI{1.65}{\gram\per\centi\meter\cubed}$ and a MgSiO\textsubscript{3}/H\textsubscript{2}O component mass fraction of \num{0.2}, resulting in a total silicate-to-ice mass ratio of \num{6.6}\footnote{However, the total amount of silicates in the model is poorly constrained by our modeling approach.}, fit the observations best as shown in Fig.~\ref{fig:water_feature_best_fit}. 

The fit to the water ice absorption feature is very good in a wavelength range of \SIrange{3.0}{3.2}{\micro\meter}, and the overall shape of the feature is well reproduced. The water ice content of \SI{5.4}{\percent} is well constrained by the depth of the feature. The systematic overestimation of flux in the red wing of the absorption feature can be attributed to the lack of the ``chemistry'' component in our model, which would increase the opacity in the corresponding wavelength range \citep[see][]{Potapov+2025}. This component contains spectral signatures primarily associated with ammonium carbamate and carbamic acid. It is required to concurrently reproduce the wing shape of the \SI{3}{\micro\meter} water ice absorption feature and some distinct absorption features in the \SIrange{6}{8}{\micro\meter} range in the compositional analysis of the d216-0939 spectrum reported by \citet{Potapov+2025}. However, the ``chemistry'' component could not be incorporated into the radiative transfer model because the required refractive indices are not available. Additionally, we observe a slight shape mismatch at wavelengths shorter than \SI{3}{\micro\meter}.

The flux decomposition by source term (see Sect.~\ref{sec:methods_comp_sim_fluxes}) shows, that the attenuated starlight, $F_\lambda^{\text{emi,}\star}$, is the dominating source term contributing \SIrange{60}{70}{\percent} to the total flux. The scattered starlight, $F_\lambda^{\text{emi,}\star}$, contributes \SIrange{30}{10}{\percent} to the integrated spectrum, decreasing significantly in the comparably narrow wavelength range of the water feature. The thermal emission of the dust, i.e., $F_\lambda^{\text{emi,dust}} + F_\lambda^{\text{sca,dust}}$, accounts for the remaining \SIrange{10}{20}{\percent} of total flux. 

Additionally to the results of the MCRT simulations shown in Figs.~\ref{fig:water_feature_F_tot_only} and \ref{fig:water_feature_best_fit}, we produced synthetic SEDs of models placing the \SI{100}{\kelvin} and \SI{150}{\kelvin} MgSiO\textsubscript{3}/H\textsubscript{2}O dust components, i.e., amorphous and crystalline water ice, respectively, into the disk model depending on the dust temperature. Specifically, we introduced a second ``snowline'' into the model, placing crystalline water ice in disk regions with temperatures of \SIrange{100}{170}{\kelvin}, and amorphous water ice in disk regions with $T < \SI{100}{\kelvin}$. Although, the second ``snowline'' temperature was chosen significantly below the expected water ice crystallization temperature of $\sim$\SI{130}{\kelvin}, we found, that the fit, especially to the position of the minimum of the water feature, is worse compared to the models containing only the \SI{150}{\kelvin} MgSiO\textsubscript{3}/H\textsubscript{2}O component. Consequently, the upper and outer disk layers contain only negligible amounts of amorphous water ice.

To assess the robustness of the results, we used model h as reference model. This model results in similar contributions of the different source terms to the total flux, but requires a slightly reduced water ice content of $\gtrsim$\SI{4}{\percent} beyond the snowline. Additionally, the intersections of the LOS with the snow- and \SI{100}{\kelvin}-line as well as the unity optical depth lines are shifted outwards by a few \si{\astronomicalunit}. However, these differences do not affect the qualitative conclusions discussed below.   

\subsubsection{Discussion}
The water ice mass fraction of $\sim$\SI{5}{\percent} beyond the snowline is low compared to the DSHARP water-ice mass fraction of \SI{20}{\percent} \citep{Birnstiel+2018}. However, the NIRSpec observations probe the upper and outer disk layers \citep{Potapov+2025}, while grain growth \citep[e.g.,][]{Birnstiel2024} and successive dust settling of larger grains \citep[e.g.,][]{Dubrulle+1995,SchraeplerHenning2004,BrunngraeberWolf2020,Sturm+2023a,Dartois+2025} may result in a vertical ice mass fraction gradient, with the ice mass increasing toward the disk midplane. Furthermore, photodesorption \citep[e.g.,][]{Hogerheijde+2011,Abrahavi+2022} could reduce the ice mass fraction in the upper disk layers.

The optical depth from the observer to the central star of the best-fit model of d216-0939 at the minimum of the water feature is $\tau^{\ \text{{ext}} \, \rightarrow \star }_{\lambda = \SI{3.1}{\micro\meter}} = \num{2.2}$. To compare the optical depth of only the water ice to previous results, we roughly estimate the optical depth of the water ice in the dust mixture as $\tau^{\ \text{{ext}} \, \rightarrow \star }_{\lambda = \SI{3.1}{\micro\meter}} - \tau^{\ \text{{ext}} \, \rightarrow \star }_{\lambda = \SI{2.8}{\micro\meter}} = \num{0.2}$, which is less than the value of $\sim$\num{0.6} found by \citet{TeradaTokunaga2012} and \citet{Potapov+2025} using a 1D LOS extinction model. Crystalline water ice in the disk atmosphere at moderate optical depths, as found for d216-0939, was also detected by \citet{SchegererWolf2010} in the young stellar object YLW 16 A. \citet{Sturm+2023b} derive a similar optical depth of \num{0.3} for the water ice feature in the upper layers of the PPD of HH 48 NE from spatially resolved JWST observations. 

The temperature structure of the best-fit disk model, shown in Fig.~\ref{fig:temperature_structure}, provides evidence for outward material transport, e.g., through a disk wind \citep[e.g.,][]{Kwan+2007,Meskini+2024,Dartois+2025}, as crystalline ice is present in disk regions, where temperatures are far below the crystallization temperature of water ice.

Dust grains of sizes $\sim$\SI{1}{\micro\meter} containing water ice are present in the upper disk layers of HH 48 NE \citep{Sturm+2023b,Sturm+2024} and d114-426 \citep{Ballering+2025}. While the ice detected in HH 48 NE is predominantly in an amorphous form, which is indicated by the round shape of the measured absorption feature, the crystalline ice in d216-0939 results in a sharp minimum of the water feature. \citet{Bergner+2026} find amorphous water ice in a sample of five other edge-on disks, making the pronounced crystalline feature detected in d216-0939 a notable outlier. As HH 48 NE is more inclined \citep[$\iota = \SI{82.3}{\degree}$;][]{Sturm+2023a} than d216-0939, a vertical crystallinity gradient of water ice in PPDs could explain the observations. Additionally, \citet{Sturm+2024} hypothesize that a dusty disk wind above the disk could be necessary to shield the water ice in the upper disk layers from photodesorption by ultraviolet irradiation and \citet{Dartois+2025} find icy dust grains at high disk altitudes and a disk wind in Tau 042021 ($\iota = \SI{87.5}{\degree}$). Assuming a similar morphology to these disks, the observations of d216-0939 could probe the extended atmosphere or a disk wind, in which crystalline ice formed close to the snowline is transported outward. To test this hypothesis, spatially resolved observations of d216-0939 in the spectral range of the water feature are necessary.

The outward material transport, successive grain growth and dust settling toward the midplane impact the chemical composition of dust available for planet formation. As the atmospheric compositions of exoplanets are used as tracers of their formation histories \citep[e.g.,][]{He+2026}, possible radial material transport in their natal disks needs to be considered. Additionally, amorphous water ice traps volatiles, such as CO and CO\textsubscript{2} \citep[e.g.,][]{Collings+2003}, which are released by volcanic desorption upon crystallization of the water ice \citep{Smith+1997,Williams+2025}. Consequently, the distribution and crystallinity of the water ice couple to the spatial distribution and chemistry of other ices. A dedicated model of the CO\textsubscript{2} ice feature of d216-0939 \citep[see][]{Potapov+2025} is necessary to assess the coupling of water ice crystallinity and distribution to the abundances of other ice species.   

\section{Conclusions}
We found a 3D model of the density structure and water ice distribution of the PPD d216-0939, that fits the SED in the near- and mid-IR, observed with JWST, as well as the spatially resolved HST flux density map in H$\alpha$. The best-fit model was analyzed to gain new insights into the spatial distribution of water ice in the PPD, which was modeled via MCRT simulations of a passively heated disk including scattering.
\begin{itemize}
\item The SED in the near- and mid-IR probes the upper disk layers, which host dust grains up to a size of $\sim$\SI{1}{\micro\meter}. This dust grain population is estimated to account for $\sim$\SI[print-unity-mantissa=false]{1e-4}{\Msun} of the total dust mass in the d216-0939 disk, resulting in a moderate optical depth from the central star to the observer of $\sim$\num{2} at the minimum of the water ice feature. We estimate the inclination angle of the edge-on disk as $\sim$\SI{78}{\degree}.
\item Direct starlight, attenuated along the LOS, accounts for up to $\sim$\SI{70}{\percent} of the total flux in the spectral region of the water ice feature. The flux contributions by scattering and thermal emission of hot dust in the PPD account for $\leq \SI{40}{\percent}$.
  \item We estimate the water ice mass fraction in the upper and outer disk layers beyond the snowline as $\sim$\SI{5}{\percent}.
  \item Crystalline water ice is present in the upper disk layers at temperatures far below the crystallization temperature of water ice, which is indirect evidence for outward material transport in the upper disk layers of the PPD.
\end{itemize}
The developed framework to model the water ice feature can be applied to JWST observations of other edge-on disks \citep[e.g.,][]{Bergner+2026}.

\begin{acknowledgements}
  We are indebted to Wolfgang Brandner for help with the HST PSFs. AP acknowledges support from the Federal Ministry for Economic Affairs and Climate Action (the German Aerospace Center project 50OR2215) and from the Deutsche Forschungsgemeinschaft (Heisenberg grant PO 1542/7-1). This research was supported in part through high-performance computing resources available at the Kiel University Computing Centre. This work made use of Astropy (\url{http://www.astropy.org}): a community-developed core Python package and an ecosystem of tools and resources for astronomy \citep{astropy:2013, astropy:2018, astropy:2022}. JSM thanks Neele Lüttkemöller for helpful discussions on the intricacies of MCRT simulations of the water ice feature. Finally, we thank the anonymous referee for the constructive comments that helped to improve the manuscript.
\end{acknowledgements}

\bibliographystyle{aa} 
\bibliography{references} 

\begin{appendix} 
  \section{Solution of the radiative transfer equation}\label{apx:radiative_transfer}
  The general RT equation describes the intensity of a PP, $I_\lambda(s)$, at position $s$ on a straight ray, i.e., the LOS. In order to write the RT equation in a compact form, we define the optical depth following the ray from $s_0$ to $s$ through a dusty medium,
  \begin{equation}
    \label{eq:optical_depth}
    \tau_\lambda(s) \equiv \int_{s_0}^{s} \varkappa^\text{ext}_\lambda(s') \varrho_\text{dust}(s') \ ds',
  \end{equation}
  via the extinction opacity, $\varkappa^\text{ext}_\lambda$, and the dust density, $\varrho_\text{dust}$. Using Eq.~\ref{eq:optical_depth}, the RT equation is written as
  \begin{equation}
    \label{eq:rt_equation}
    \frac{d I_\lambda}{d \tau_\lambda} = -I_\lambda(s) + S_\lambda(s)
  \end{equation}
  with the nonlinear source term, $S_\lambda(s)$, including thermal dust emission and scattering into the LOS of the ray \citep[e.g.,][]{RybickiLightman1979}. To find a solution of the RT equation, it is expedient to split the numerical solution into two steps: First, we find the thermal equilibrium of the dust irradiated by a radiation source with an intensity field, $S_\lambda^{\text{emi,}\star}(\vec{r})$, employing the \citet{BjorkmanWood2001} algorithm. Second, the solution of the RT equation along an arbitrary ray is computed by splitting the source term into the following contributions:
  \[
  \begin{array}{lp{0.8\linewidth}}
    S_\lambda^{\text{emi,}\star} & Emission of the radiation source; \\
    S_\lambda^\text{emi,dust} & Thermal emission of the dust along the ray; \\
    S_\lambda^{\text{sca,}\star} & Scattered starlight into the LOS; \\
    S_\lambda^\text{sca,dust} & Thermal emission of the dust scattered off of the dust into the LOS (``self-scattering'').
  \end{array}
  \]
  Considering an initial intensity of $I_\lambda(s_0)$, the solution of the RT equation is
  \begin{equation}
          \tiny
    \begin{aligned}
      I_\lambda(s) =& \underbrace{I_\lambda(s_0) e^{-\tau_\lambda(s)} + \int_{s_0}^{s} \left( S_\lambda^{\text{emi,}\star}(s') + S_\lambda^\text{emi,dust}(s') \right) e^{-\big(\tau_\lambda(s)- \tau_\lambda(s')\big)} \ ds'}_{\text{LOS emission and attenuation by extinction}} \\
      &+ \int_{s_0}^{s} \underbrace{\frac{1}{4\pi} \iint S_\lambda^{\text{sca,}\star}(s',\Omega) + S_\lambda^{\text{sca,dust}}(s',\Omega) \ d\Omega}_{\text{radiation scattered into the LOS}} e^{-\big(\tau_\lambda(s)- \tau_\lambda(s')\big)} \ d s'. 
      \end{aligned}
  \end{equation}
  The integral of scattered radiation into the LOS over the full solid angle, $\Omega$, is estimated numerically using the Monte Carlo technique \citep[e.g.,][]{Wolf+1999,Pinte+2006,Steinacker+2013}.\newline

  \section{Validation and performance of the spectral importance sampling method}
  \label{apx:gespenst}
  The spectral importance sampling method, described in Sect.~\ref{sec:spectral_importance_sampling}, is implemented in the MCRT code \texttt{gespenst}\footnote{\url{https://github.com/jammartin/gespenst}}, which was validated against \texttt{POLARIS} \citep{Reissl+2016,Reissl+2018} simulation results of scattering and dust emission of an accretion disk model.

\begin{figure}
  \centering
  \includegraphics[width=\columnwidth]{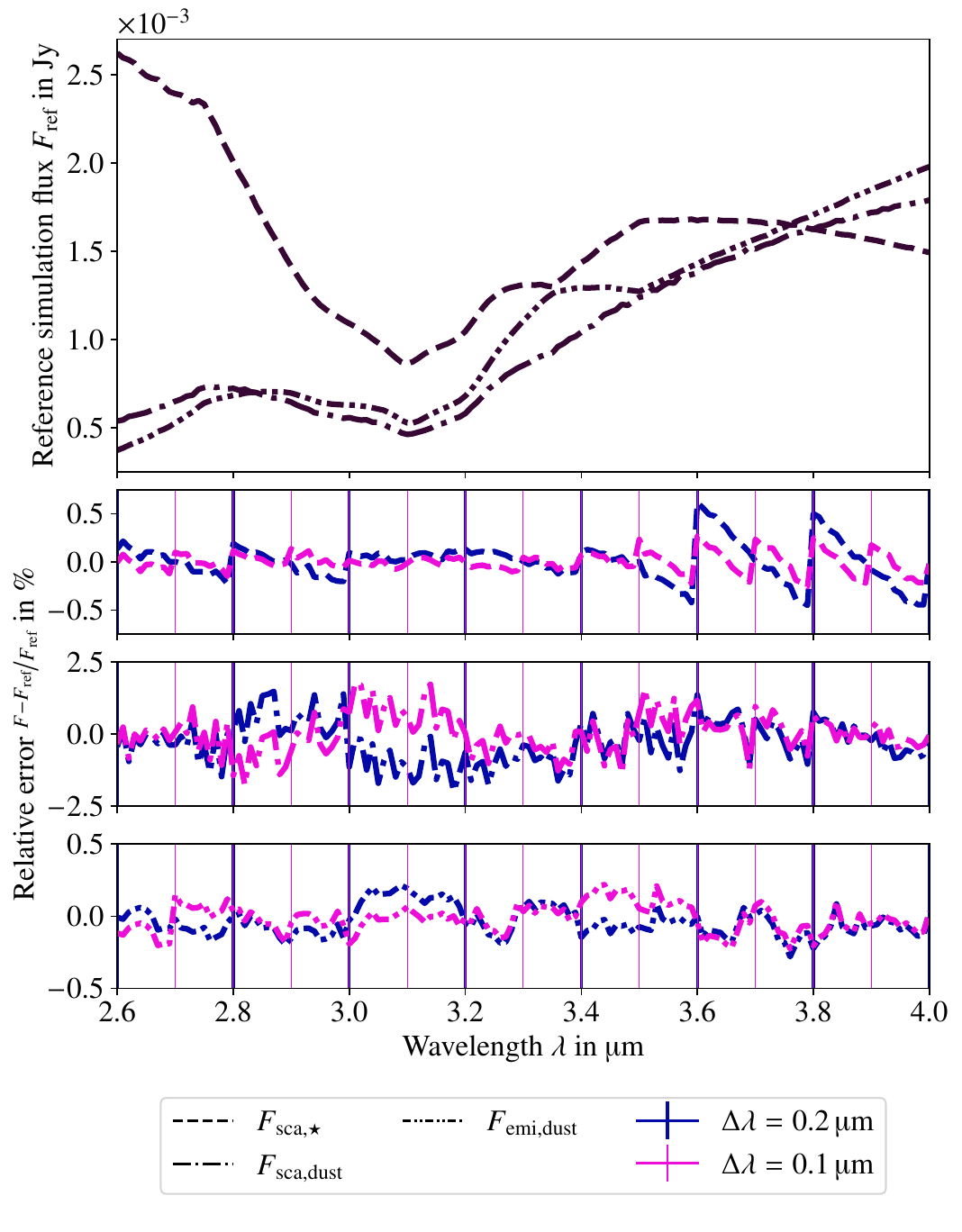}
  \caption{\emph{Top row:} Scattered starlight, dust emission, and self-scattering computed via wavelength-by-wavelength MCRT simulations. \emph{Bottom rows:} Relative errors of polychromatic MCRT simulations with varying wavelength binwidth with respect to the reference simulation results. The borders of the wavelength bins, in which spectral importance sampling is utilized, are indicated by vertical lines.}
  \label{fig:spectral_importance_sampling}
\end{figure}

  To demonstrate the performance and accuracy of the spectral import sampling method, we ran RT simulations with wavelength binwidths of $\Delta \lambda \in \{ \num{0}, \num{0.1}, \num{0.2} \} \, \si{\micro\meter}$ and 141 equally spaced wavelengths in the water feature wavelength range, \SIrange{2.6}{4.0}{\micro\meter}, of the reference model (see Sect.~\ref{sec:continuum_baseline_disk_model}) using dust mixture \# 2 (see Table \ref{tab:ice_mixtures}) to incorporate water ice into the model as described in Sect.~\ref{sec:snowline_model}. The relative errors, with respect to the reference simulations with $\Delta \lambda = \SI{0}{\micro\meter}$, i.e., classical, wavelength-by-wavelength MCRT simulations, of scattered starlight, thermal dust emission, and self-scattering are shown in Fig.~\ref{fig:spectral_importance_sampling}. The discontinuities of the relative error of the scattering of starlight simulations at the wavelength bin borders indicate the systematic error introduced by the spectral importance sampling method, which is $\lesssim$\SI{0.5}{\percent} and decreases for smaller spectral binwidths. Note that this error would vanish for an infinite amount of PPs. The relative error of the dust emission and self-scattering simulations is dominated by the statistical error intrinsic to the Monte Carlo approach. The corresponding computation times and numbers of PPs used in the simulations, $N_\text{PP}$, are listed in Table \ref{tab:spectral_importance_sampling}. Conclusively, the spectral importance sampling method reduces the computation time of MCRT simulations significantly at the cost of introducing only small systematic errors.
  \begin{table}
\caption{Overview of MCRT simulations performed to validate the spectral importance sampling method.}             
\label{tab:spectral_importance_sampling}      
\centering
\begin{tabular}{cccr}
  \toprule
  & $N_\text{PP}$ & $\Delta \lambda$ & Computation time \\
  \midrule
  &  & -- & \SI{19}{\hour} \SI{10}{\minute} \\
  Scattering of starlight & \num[print-unity-mantissa=false]{1e8} & \SI{0.1}{\micro\meter} & \SI{5}{\hour} \SI{14}{\minute} \\
  &  & \SI{0.2}{\micro\meter} & \SI{3}{\hour} \SI{27}{\minute} \\
  \midrule
  &  & -- & \SI{1}{\hour} \SI{16}{\minute} \\
  Dust emission & \num[print-unity-mantissa=false]{1e8} & \SI{0.1}{\micro\meter} & \SI{28}{\minute} \\
  &  & \SI{0.2}{\micro\meter} & \SI{19}{\minute} \\
  \midrule
  &  & -- & \SI{5}{\day} \SI{17}{\hour} \SI{49}{\minute} \\
  Self-scattering & \num[print-unity-mantissa=false]{1e7} & \SI{0.1}{\micro\meter} & \SI{2}{\day}~~~\SI{4}{\hour} \SI{11}{\minute} \\
  &  & \SI{0.2}{\micro\meter} & \SI{1}{\day}~~~\SI{8}{\hour} \SI{54}{\minute} \\
  \bottomrule
\end{tabular}
\tablefoot{The simulations were performed on a machine equipped with two AMD EPYC\texttrademark \ 7643 processors with 48 cores each, utilizing 192 threads in total.}
\end{table}
  
  \section{Mass normalization of the disk model}
  \label{apx:mass_normalization}
  The total gas mass,
  \begin{equation}
    \label{eq:total_mass}
    \tiny
    \begin{aligned}
    M_\text{disk} = \int_0^{2\pi} \int_{R_\text{in}}^{R_\text{max}} \int_{-\sqrt{R_\text{max}^2-r^2}}^{\sqrt{R_\text{max}^2-r^2}} & \, \varrho_\text{disk}(r, z) \, dz \, r \, dr \, d\varphi \\
    = (2 \pi)^{\nicefrac{3}{2}} \varrho_\text{ref} \, h_\text{ref}\, R_\text{ref} \int_{R_\text{in}}^{R_\text{max}} & \left( \frac{R_\text{ref}}{r} \right)^{\alpha-\beta-1} \exp\left[ - \left( \frac{R_\text{out}}{r} \right)^{\alpha-\beta-2} \right] \\
    & \erf\left[\frac{\sqrt{R_\text{max}^2-r^2}}{\sqrt{2}h_\text{ref}} \left(\frac{R_\text{ref}}{r}\right)^\beta\right]  \, dr ,
    \end{aligned}
  \end{equation}
  in the disk model (see Sect.~\ref{sec:disk_structure}) is evaluated numerically. The reference density, $\varrho_\text{ref}$, yielding a certain disk mass, $M_\text{disk}$, is found via Newton-Raphson iteration.
 
  \section{Computation of dust mixture bulk densities}\label{apx:effective_medium_icy_dust}
  We computed the bulk densities of the MgSiO\textsubscript{3}/H\textsubscript{2}O material, 
\begin{equation}
  \rho_\text{sil/ice} = \frac{\rho_\text{sil} \rho_\text{ice} \left( 1+\nicefrac{m_\text{sil}}{m_\text{ice}} \right) }{\rho_\text{sil} + \rho_\text{ice} \nicefrac{m_\text{sil}}{m_\text{ice}} },
\end{equation}
using the silicate-to-ice mass ratio $\nicefrac{m_\text{sil}}{m_\text{ice}} = \num{2.7}$ \citep{Potapov+2021}, $\rho_\text{sil} = \SI{2.5}{\gram\per\centi\meter\cubed}$ \citep[MgSiO\textsubscript{3};][]{Potapov+2018}, and water ice densities $\rho_\text{ice}(\SI{100}{\kelvin}) = \SI{0.94}{\gram\per\centi\meter\cubed}$ and $\rho_\text{ice}(\SI{150}{\kelvin}) = \SI{0.93}{\gram\per\centi\meter\cubed}$ \citep{LaSpisa+2001}.

\begin{figure}
  \centering
  \includegraphics[width=.7\columnwidth]{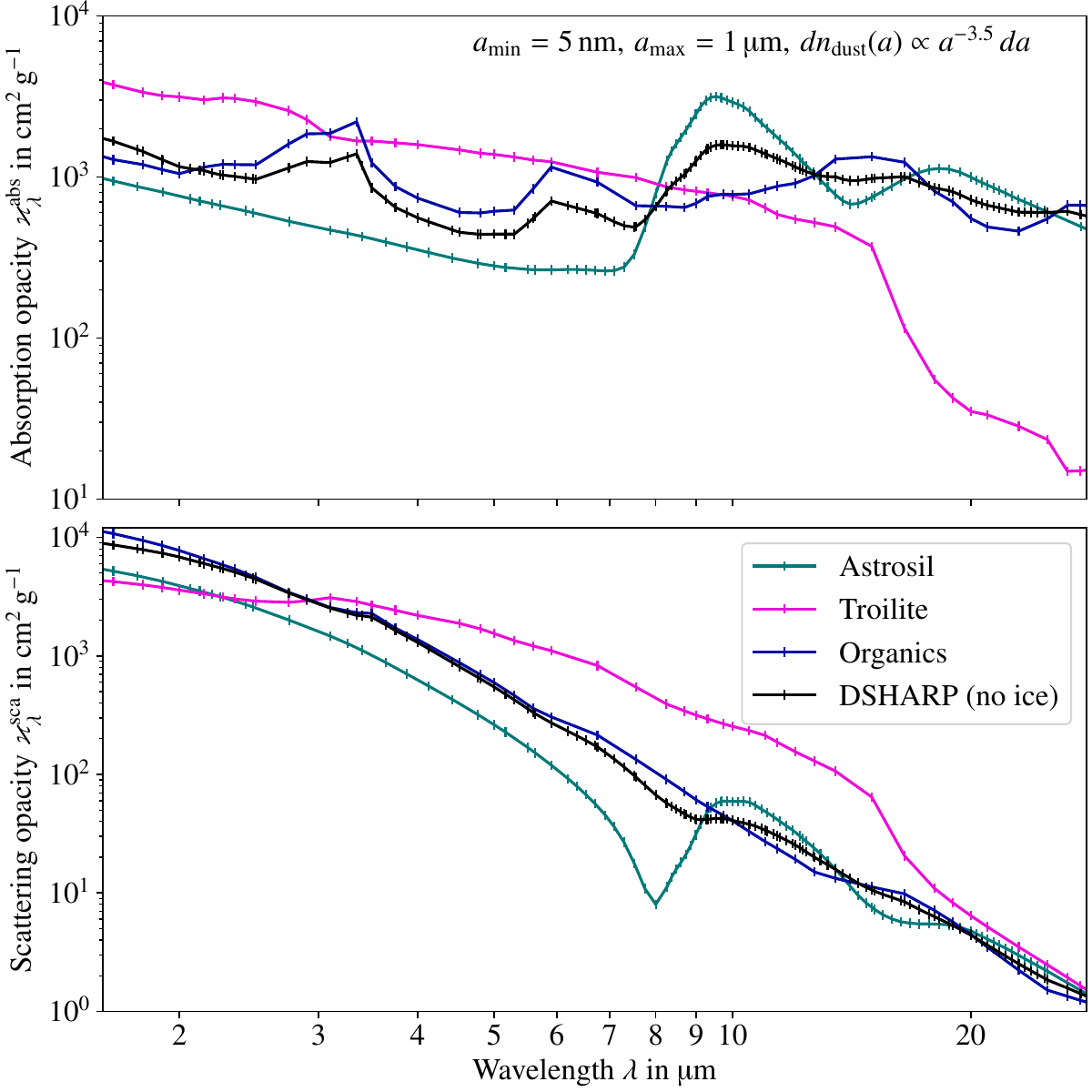}
  \caption{Opacities of the DSHAPR (no ice) dust mixture and its
    constituent dust components in the JWST wavelength range.}
  \label{fig:dsharp_decomposition}
\end{figure}

The densities of effective medium dust mixtures,
\begin{equation}
  \rho_\text{mix} = \left( \sum_c \frac{f_c}{\rho_c} \right)^{-1}, 
\end{equation}
were computed using the mass fractions $f_c$ and bulk densities $\rho_c$ of the dust components, $c$, constituting the mixture. The opacities of the DSHARP (no ice) dust mixture (see Sect.~\ref{sec:dust_properties}) are shown in Fig.~\ref{fig:dsharp_decomposition}.  

\FloatBarrier
\section{Best-fit candidate models}

\begin{table}
\caption{Best-fit candidate models.}             
\label{tab:best-fit_continuum_models}      
\centering
  \tiny
\begin{tabular}{cccccc}
  \toprule       
  Model & $L_\star$ & $M_\text{disk}$ & $h_\text{ref}$ & $R_\text{out}$ & $\iota$ \\
  & $(\si{\Lsun})$ & $(\si{\Msun})$ & $(\si{\astronomicalunit})$ & $(\si{\astronomicalunit})$ & $(\si{\degree})$ \\ 
  \midrule
  a  & \num{1.25} & \num{2e-3} & \num{15.0} & \num{500} & \num{72.5} \\ 
  b  & \num{1.25} & \num{8e-4} & \num{17.5} & \num{300} & \num{72.5} \\ 
  c  & \num{1.25} & \num{2e-3} & \num{15.0} & \num{400} & \num{72.5} \\ 
  d  & \num{1.25} & \num{4e-4} & \num{17.5} & \num{400} & \num{75.0} \\ 
  e  & \num{1.25} & \num{8e-4} & \num{17.5} & \num{400} & \num{72.5} \\ 
  f  & \num{1.25} & \num{4e-4} & \num{17.5} & \num{500} & \num{75.0} \\ 
  g  & \num{1.00} & \num{8e-4} & \num{17.5} & \num{500} & \num{72.5} \\ 
  h  & \num{1.00} & \num{4e-4} & \num{17.5} & \num{200} & \num{72.5} \\ 
  i  & \num{1.00} & \num{8e-4} & \num{17.5} & \num{400} & \num{72.5} \\ 
  j  & \num{1.25} & \num{1e-4} & \num{17.5} & \num{300} & \num{77.5} \\ 
  k  & \num{1.25} & \num{1e-4} & \num{17.5} & \num{400} & \num{77.5} \\ 
  l  & \num{1.25} & \num{8e-4} & \num{15.0} & \num{500} & \num{75.0} \\ 
\bottomrule                  
\end{tabular}
\tablefoot{The flaring exponent and maximum grain size are $\beta=\num{1.250}$ and $a_\text{max} = \SI{1}{\micro\meter}$, respectively.}
\end{table}
  \begin{figure}[h]
  \centering
  \includegraphics[width=.9\columnwidth]{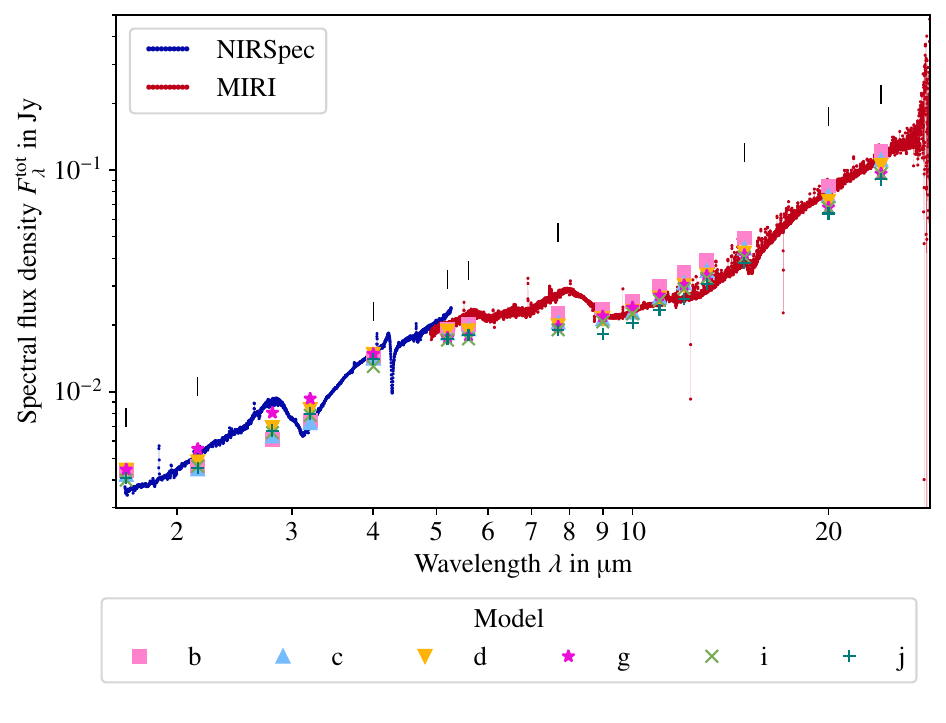}
  \caption{Comparison of the observed SED to the well-fitting disk models, that were not further considered (see Sect.~\ref{sec:continuum_baseline_disk_model}). The free parameter values of the models are listed in Table \ref{tab:best-fit_continuum_models}. The wavelengths $\lambda_\text{cont}$ are marked by vertical black lines above the data.}
  \label{fig:best-fit_candidates_deselected}
\end{figure}

\end{appendix}

\end{document}